# Design study for an efficient semiconductor quantum light source operating in the telecom C-band based on an electrically-driven circular Bragg grating

Andrea Barbiero[1,2], Jan Huwer[1], Joanna Skiba-Szymanska[1], Tina Müller[1], R. Mark Stevenson[1] and Andrew J. Shields[1]

[1] *Toshiba Europe Limited, Cambridge Research Laboratory, 208 Science Park, Milton Road, Cambridge, CB4 0GZ, UK*
[2] *Department of Physics and Astronomy, University of Sheffield, Hounsfield Road, Sheffield, S3 7RH, UK*

Corresponding author: andrea.barbiero@crl.toshiba.co.uk

**Abstract:** The development of efficient sources of single photons and entangled photon pairs emitting in the low-loss wavelength region around 1550 nm is crucial for long-distance quantum communication. Moreover, direct fiber coupling and electrical carrier injection are highly desirable for deployment in compact and user-friendly systems integrated with the existing fiber infrastructure. Here we present a detailed design study of circular Bragg gratings fabricated in InP slabs and operating in the telecom C-band. These devices enable the simultaneous enhancement of the X and XX spectral lines, with collection efficiency in numerical aperture 0.65 close to 90% for the wavelength range 1520 - 1580 nm and Purcell factor up to 15. We also investigate the coupling into single mode fiber, which exceeds 70% in UHNA4. Finally, we propose a modified device design directly compatible with electrical carrier injection, reporting Purcell factors up to 20 and collection efficiency in in numerical aperture 0.65 close to 70% for the whole telecom C-band.

## 1. Introduction

The generation of photons with high purity and indistinguishability is an essential building block for emerging photonic quantum technologies, such as quantum key distribution, quantum communication networks and quantum computing, that started changing the way information is transferred and processed [1,2]. The last few years have seen important progress in the development of efficient non-classical light sources based on the combination of semiconductor quantum dots (QDs) with vertically emitting photonic structures, including nanowires [3–6], cavities [7–12] or microlenses [13–16]. However, the most mature devices are based on InAs/GaAs QDs and operate around 900 nm, which prevents their integration with the standard optical fiber infrastructure due to the strong attenuation (>1dB/km) at those wavelengths. Recent efforts have focused on devices operating in the telecom O-band [17–19] where the attenuation is only 0.35dB/km, but for long-haul quantum communication and quantum network schemes it is desirable to exploit the optimal low-loss window of silica fibers in the telecom C-band, where attenuation can be as little as 0.12dB/km.

Quantum light sources based on InAs QDs emitting directly around 1550 nm have been developed using two different approaches: they can be either grown on GaAs-based materials by employing metamorphic buffer layers for strain relaxation [20] or on InP substrates, which have smaller lattice mismatch with respect to the QD material and therefore allow for the deposition of less strained islands [21,22]. Between those two methods, QDs grown on InP have shown superior performances in the coherence of the emission [23–25], which is a crucial figure of merit for efficient quantum communication protocols. However, those results have been achieved with simple structures such as planar DBR microcavities and the investigation of more complex devices is highly desirable for the development of efficient and competitive quantum light sources based on the InP material system.

In this context, hybrid circular Bragg gratings (CBGs) [26–29] recently emerged as an attractive concept thanks to a new fabrication process for the integration of a backside gold mirror, which improves the robustness and the reproducibility of the nanostructure as well as the optical performances compared to the original suspended version and subsequent variations [30–34]. The first experiments with hybrid CBGs showed a unique combination of high brightness, entanglement fidelity and photon indistinguishability [27,28] and generated much interest in this solution, which led to further works such as a design optimization and the fabrication of the first device for the telecom O-band [35,36] and the demonstration of a strain-tunable device [37]. Nonetheless, a design study of InP CBGs operating in the telecom C-band is still missing. Moreover, the original layout of hybrid CBGs is not compatible with electrical carrier injection because the central disk hosting the emitter is completely isolated from the rest of the semiconductor slab. A recent work [38] tried to overcome this problem by combining a CBG with a doped bottom DBR mirror, achieving an efficiency of 24%. The more advanced concept of incorporating a CBG into a vertical hybrid passive cavity [39] could in principle increase the efficiency up to 79% for a 8 nm bandwidth, but appears very challenging to implement in practice because it requires etching narrow trenches with straight sidewalls and aspect ratios around 10:1. Overall, none of the variations proposed so far can provide the combination of moderate Purcell enhancement and high extraction efficiency in a broad range that makes the original design so attractive.

In this paper we present a detailed theoretical investigation of circular Bragg gratings fabricated in InP slabs and operating in the telecom C-band. Using FEM numerical simulations, we optimize the choice of the design parameters to achieve the best combination of Purcell factor, bandwidth and collection efficiency: we report an efficiency in numerical aperture NA = 0.65 close to 90% for the whole wavelength range 1520 - 1580 nm together with a cavity mode exhibiting FWHM = 16 nm and Purcell factor up to 15. We also determine the best strategy for direct coupling of the emitted photons into a single mode (SM) fiber, which is essential for applications that require compact and deployable quantum light sources that can be used outside of a research laboratory. Furthermore, we propose a modified device design which is directly compatible with electrical carrier injection, with Purcell factor up to 20 and collection efficiency in NA = 0.65 close to 70% for the whole telecom C-band.

## 2. Methods

For this design study we employ 3D FEM simulations to solve the wave equation in the frequency domain. Literature values are used for the refractive indices of all materials [40–42]. The simulation domain has cylindric symmetry along the z axis and is meshed with seven tetrahedral elements per effective wavelength. The computational region is enclosed within perfectly matched layers (PMLs), i.e. artificial strongly absorbing layers that simulate open boundaries and avoid undesired reflections of the emitted radiation. The PMLs are meshed with a rectangular swept mesh. Finally, an electric point dipole source oriented along the y axis is placed in the center of the device (x = 0, y = 0) at half of the slab thickness, in order to simulate the single photon emission from a single quantum dot.

We calculate the Purcell factor by monitoring the power outflow through the boundaries of the model: $F_p$ is obtained as the ratio between the power flow generated by the dipole source in the CBG and in semi-infinite InP. In order to evaluate the efficiency of the devices and following the convention introduced in previous reports [26,35], we define the dipole extraction efficiency (DEE) as the percentage of radiation emitted in the air domain, the far-field efficiency (FFE) in a certain NA as the far-field intensity integrated on such NA divided by the far-field intensity in the upper air hemisphere and the dipole collection efficiency (DCE) as the product of the previous two quantities.

## 3. Results

### 3.1 Optimization of Purcell factor and collection efficiency

The aim of the first part of this study is to find the best combination of Purcell factor, bandwidth and collection efficiency for a CBG operating in the telecom C-band. In Fig. 1a and Fig. 1b we show a sketch and a schematic cross-section of the device investigated, which consists of an InP disk surrounded by a grating, an insulating layer of $SiO_2$ and a backside gold mirror. The design parameters initially studied are the thickness of the different layers ($t_{InP}$, $t_{SiO2}$, $t_{Au}$), the radius $R$ of the central disk, the number $N$ of concentric InP rings, their lattice constant (i.e. periodicity) $a$ and the width of the trenches $w$. The choice of the initial group of parameters [$P_0$] is based on previous works at shorter wavelengths [30,31]: in particular, we set the initial thickness of InP to $t_{InP} = \lambda_{eff}/2$ (where $\lambda_{eff}$ is the effective wavelength within the material) to have only a single TE mode supported by the slab, the grating period to $a = \lambda/n_{TE}$ to satisfy the second-order Bragg condition and the radius of the central disk to $R = \lambda_{eff}$. Moreover, an initial thickness of the oxide layer $t_{SiO2} = 270$ nm guarantees that the vertical separation between the center of the InP disk and the Au mirror corresponds to $\lambda_{eff}/2$. We arbitrarily set starting values for the width of the trenches and the number of rings to $w = a/2$, $N = 10$ and $t_{Au} = 100$ nm.

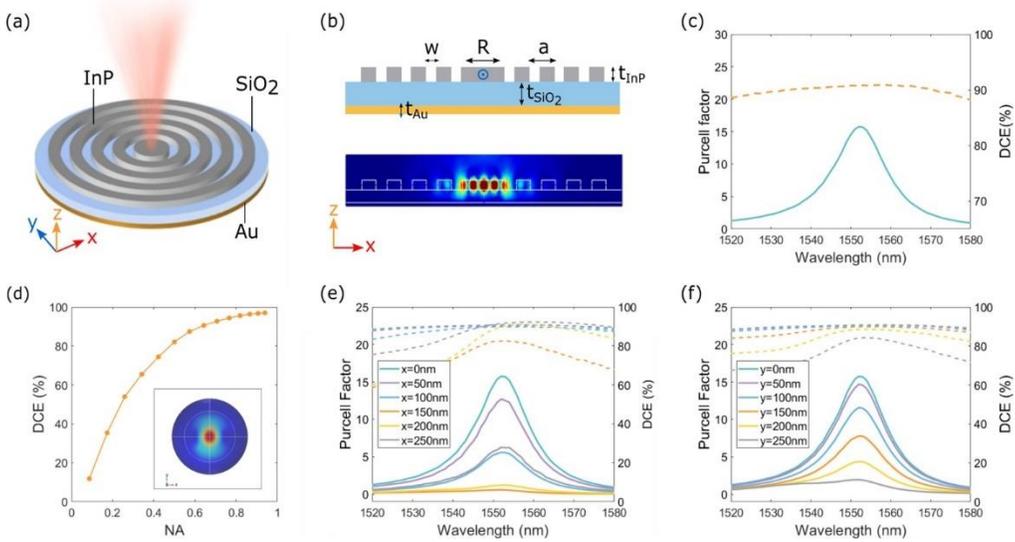

**Fig. 1** (a) 3D illustration of the hybrid CBG with light emission from a QD embedded in the central disk. The device consists of an InP disk and grating, an insulating layer of $SiO_2$ and a backside gold mirror. (b) Schematic cross-section of the CBG in the xz plane and corresponding electric field profile $|E|^2$ of the cavity mode confined in the central InP disk. (c) Simulated performances of the device, with Purcell factor (teal) and collection efficiency (DCE) in NA = 0.65 (orange) as a function of wavelength. (d) DCE of the optimized deviced calculated for $\lambda = 1552$ nm as a function of the collecting NA. The inset shows the corresponding far-field intensity distribution projected on a sphere, with NA = 0.65 represented by the inner circle. (e, f) Effect of the dipole displacement on the performances of the CBG, with Purcell factor and DCE in NA = 0.65 as a function of wavelength calculated for different positions of the dipole along (e) the x axis and (f) the y axis.

The optimization process consists of three main steps: we start with a 1D InP grating (i.e. a radial section of the CBG) sitting on $SiO_2$. A waveguide TE mode is launched in the slab in order to verify that such a grating can provide high reflectivity and out-of-plane scattering in the telecom C-band [27]. Then, in a first series of 3D simulations the thickness of the different layers and radius of the central disk are varied trying to maximize the Purcell factor and DCE around λ = 1550 nm. This produces a preliminary set of output parameters [$P_1$]. Next, in a second series of 3D simulations we investigate finer variations to [$P_1$] and monitor the performances of the CBG in the wavelength range between 1520 nm and 1580 nm. It is worth noting that this last step does not aim at maximizing the performances at a fixed wavelength. Instead, the goal is to find the best compromise between Purcell factor, bandwidth and DCE for a versatile device that can enhance QD transitions in the whole telecom C-band without major design adjustments.

The final set of design parameters [$P_2$] produced by the optimization process is: $t_{InP}$ = 280 nm, $t_{SiO2}$ = 360 nm, $R$ = 660 nm, $a$ = 785 nm, $w$ = 350 nm, $N$ = 4. The performances of the optimized device are summarized in Fig. 1c, which shows that collection efficiencies close to 90% in NA = 0.65 are expected for the whole telecom C-band, together with a cavity mode exhibiting FWHM = 16 nm and Purcell factor up to 15. It is important to underline that a higher $F_p$ could be easily achieved by increasing the Q factor of the cavity (Fig. S1). Nevertheless, the less selective cavity design of Fig. 1c guarantees at the same time moderate Purcell enhancement and efficient photon collection in a broad range of wavelengths encompassing both the X and XX spectral lines, typically separated by 5 nm in InAs/InP QDs [24,43]. In fact, according to recent works on Droplet Epitaxy (DE) InAs/InP QDs [24,25], the enhancement of radiative emission rate $1/T_1$ generated by a cavity mode with $F_p$ = 10 would be sufficient to bring both transitions to the Fourier limit.

It is also worth noting that the high directionality of the emission, which is clearly observable both in the near-field and far-field intensity distribution, guarantees high collection efficiency even in a smaller NA, with DCE exceeding to 80% for NA = 0.5 (Fig. 1d).

When a CBG is fabricated, it is unlikely to find a quantum dot in the ideal central position (x = 0, y = 0) and, for an accurate estimate of the behaviour of a real device, it is important to simulate this imperfection by moving the dipole source away from the center. Therefore, we ran a further set of simulations maintaining the original orientation of the dipole, but placing it at different positions along the x and y axes. When the dipole is moved along the x axis (Fig. 1e) the DCE stays almost constant up to x = 100 nm, then it drops to 80% around resonance with a more pronounced wavelength dependence. On the other hand, the Purcell factor decreases more rapidly, giving $F_p$ = 5 for x = 100 nm. It is worth noting that after reaching a minimum for x = 150 nm, $F_p$ increases again while approaching the next maximum of $|E|^2$ in the central disk. Nevertheless, a small displacement up to 50 nm is desirable in order to keep $F_p$ closer to the original value of 15 and maintain a flat and broad DCE.

A similar behaviour of the DCE is observed for a displacement along the y axis (Fig. 1f), while the reduction of the Purcell factor is less steep. The simulated asymmetry of $F_p$ is consistent with previous reports [35] and originates from the specific orientation of the dipole source used to excite the structure. Since an InAs QD would either show a superposition of both x and y polarizations (neutral excitons) or circularly polarized emission (charged excitons), we expect the near-field profile to be circularly symmetric. Moreover, thanks to the recent progress in marker-based fabrication techniques [44], where first a PL imaging system is used to localize a single QD with respect to alignment features [45,46] and then the device is fabricated around the target emitter by EBL, deterministic fabrication of CBGs with high lateral accuracy (< 30 nm) has already been achieved [32,33]. For such a small dislocation our simulations do not predict a relevant decay of the optical performances, therefore the fabrication of the proposed InP CBG is expected to show good robustness against a lateral displacement of the QD emitter.

Finally, we would like to emphasize that, as reported in Fig. S2 where we analyze the individual design parameters and their influence on the performances of the device, small variations of $R$ and $a$ can be used for a fine adjustment of the resonance wavelength of the CBG with minimal impact on the collection efficiency. This could be exploited to compensate for a potential shift in the cavity mode of a real device due to slightly inaccurate values of the refractive indices or imperfections in the fabrication process.

## 3.2 Direct coupling into SM fiber

For the development of compact and user-friendly QD-based sources intended to be used outside the lab environment, direct fiber coupling of the emitted photons is most desirable. In fact, having an optical fiber directly attached to the devices would reduce the size and complexity of the system and guarantee high alignment robustness [47–50]. Therefore, our next set of simulations investigates the optimal coupling into a SM fiber: using a field overlap method [51] to calculate the mode coupling efficiency (MCE), we select the best fiber to achieve high coupling efficiency and present a further optimization of the CBG aiming to maximize this parameter, which is the most crucial figure of merit for a bright plug-and-play quantum light source. In our study we consider 4 types of commercial SM fibers (Nufern), with specifications at $\lambda = 1550$ nm reported in Table 1. We assume that the fiber is positioned on top of the device at a distance $z$, perfectly aligned in the axial direction and with vacuum ($n = 1$) in the gap (Fig. 2a).

**Table 1.** Specifications of the SM fibers investigated in our numerical simulations provided by Thorlabs Inc.

| Fiber | n core | d core (um) | n cladding | d cladding (um) | NA |
|---|---|---|---|---|---|
| 980HP | 1.4507 | 3.6 | 1.4457 | 125 | 0.12 |
| UHNA1 | 1.4709 | 2.5 | 1.444 | 125 | 0.28 |
| UHNA3 | 1.4858 | 1.8 | 1.444 | 125 | 0.35 |
| UHNA4 | 1.4858 | 2.2 | 1.444 | 125 | 0.35 |

In Fig. 2b we report the MCE as a function of the distance $z$ between the fiber and the device calculated at $\lambda = 1550$ nm for the four different SM fibers described above. Only a small fraction (up to 22.8%) of the light emitted by the CBG is coupled into the 980HP, while using the UHNA1 fiber results in a remarkable improvement of the MCE up to 64.4%, with a decay for $z > 2$ µm. Even higher efficiencies can be achieved with the more extreme versions UHNA3 and UHNA4. All the curves also exhibit moderate oscillations caused by Fabry-Perot interference in the gap region [35], with the maximum coupling efficiency at $z = 800$ nm. We attribute the superior performances of the UHNA4 fiber to its high NA and small mode field diameter, that is very well matched with the spatial distribution of the CBG emission (Fig. 2c). It is important to mention that UHNA fibers can be fusion spliced directly into standard SMF28 with low losses ($\leq 0.15$ dB) [52], which guarantees easy compatibility with standard fiber networks [53].

Since small variations of the trench width have a negligible effect on the resonance wavelength and do not reduce significantly the Purcell factor (Fig. S2), we analyze this parameter further and report the wavelength dependent MCE in UHNA4 fiber for different values of $w$. Fig. 2d shows that etching trenches with a width of 370 nm provides even better coupling, with MCE above 71% at 1550 nm for $z = 1.6$ µm.

The fiber alignment in practical applications could be achieved by taking advantage of interferometric methods [54]. Alternatively, one could employ 3D printed fiber–holders, which have been successfully shown to provide accurate positioning and sustain cryogenic cycling without signs of degradation [55,56]. Finally, it is worth noting that introducing an adhesive in the gap region for permanent coupling may affect the optimal distance $z$ and the maximum MCE [35].

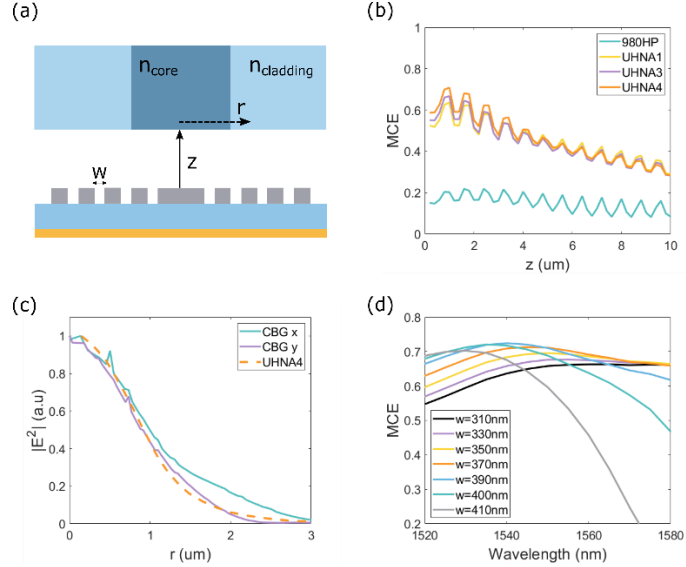

**Fig. 2.** (a) Schematic of the model created to investigate the coupling efficiency, with a SM fiber at a distance $z$ from the CBG (b) Mode coupling efficiency (MCE) as a function of the vertical distance $z$ calculated at resonance. (c) Comparison between the mode profile of an UHNA4 fiber (dashed line) and the spatial profile of the near-field emission from the CBG along the x and y axis (solid lines) calculated for $z = 1.6$ µm. (d) Coupling efficiency into UHNA4 fiber as a function of wavelength for different widths $w$ of the trenches. The efficiency is calculated for $z = 1.6$ µm.

### 3.3 Electrically driven circular Bragg grating

The structure of the hybrid CBG presented above does not permit any direct electrical control of the QD emission: in fact, the fully etched circular trenches isolate the central disk from any type of metal contact placed outside the device. Therefore, in the final set of simulations we investigate a modified design compatible with electrical carrier injection. We propose a p-i-n slab structure with increased $t_{InP} = 400$ nm (Fig. 3a), which includes two 50 nm-thick doped layers. As shown in Fig. 3b, we also avoid etching fully circular trenches and leave four semiconductor bridges of width $w_b$ that connect the central disk to the n-type and p-type contacts placed outside the device [34]. The initial value of $w_b$ is set to 200 nm, which is easy to achieve with current lithographic techniques. Since the introduction of the semiconductor bridges breaks the circular symmetry of the CBG, we expect the optical performances to be influenced by the orientation of the dipole source with respect to the bridges quantified by the relative angle α. For the purpose of this study, we decided to focus on the two limit cases of $α = 0°$ and $α = 45°$, that represent two individual photon-emission events with defined polarization.

The final set of design parameters produced by the numerical optimization is: $t_{SiO2} = 280$ nm, $R = 610$ nm, $a = 740$ nm, $w = 350$ nm, $N = 4$. The performances of the optimized device for the two different limit cases are summarized in Fig. 3c and Fig. 3d, which show that collection efficiencies close to 70% in NA = 0.65 are expected for the whole telecom C-band. In both cases, the cavity resonance is nicely confined in the central disk, exhibiting Purcell factor up to 20. The higher value of $F_p$ at resonance compared to the optically pumped version is the consequence of a more selective cavity design with FWHM = 10 nm.

Noteworthy, the orientation of the dipole with respect to the bridges has a negligible effect on the DCE and causes only a small red shift (~3 nm) of the cavity mode. Considering the large bandwidth of the device, this does not appear to be an obstacle to the simultaneous enhancement of the X and XX transitions. Given the typical radiative lifetimes ~1.5 ns of the excitonic states in DE InAs/InP QDs [25], these results pave the way for the realization of an efficient and GHz-clocked quantum light source.

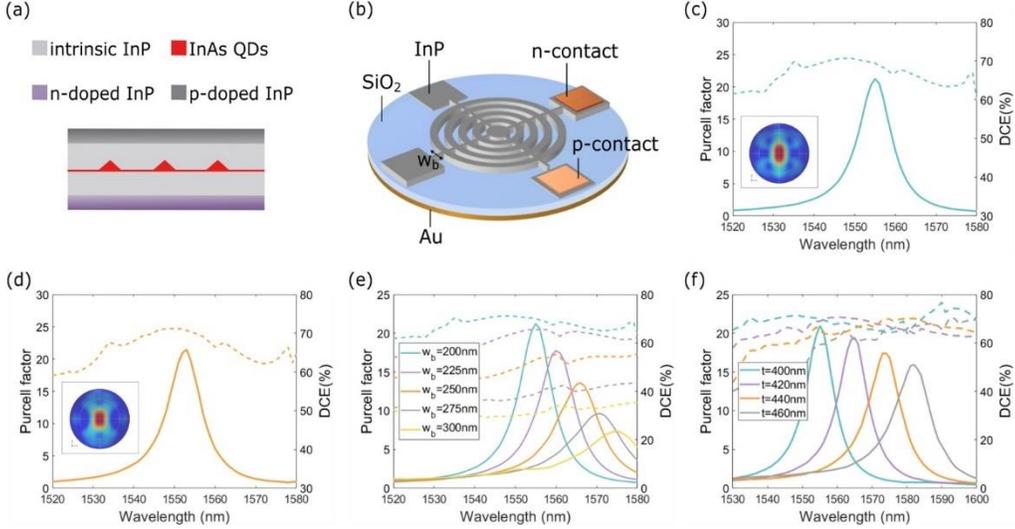

**Fig. 3.** (a) Schematic of the InP p-i-n slab structure as grown, including two 50 nm-thick n-doped and p-doped layers. (b) 3D illustration of the electrically controllable CBG. The metal contacts are deposited on two of the four squares outside the CBG, which are not in scale in this drawing. (c,d) Simulated Purcell factor (solid line) and collection efficiency (DCE) in NA = 0.65 (dashed line) as a function of wavelength for (c) $α = 0°$ and (d) $α = 45°$. The insets show the corresponding far-field intensity distribution projected on a sphere, with NA = 0.65 represented by the inner circle. (e) Influence of the bridges on the performances of the device: Purcell factor (solid lines) and DCE in NA = 0.65 (dashed lines) as a function of wavelength for five different values of $w_b$. (f) Influence of the InP thickness on the performances of the device: Purcell factor (solid lines) and DCE in NA = 0.65 (dashed lines) as a function of wavelength for five different values of $t_{InP}$.

As reported in Fig. 3e, the performances of the CBG are significantly affected by the width of the InP bridges, with a constant descent of both $F_p$ and DCE for increased $w_b$ that we attribute to a weaker confinement of the cavity mode caused by the intermittent grating. On the other hand, the device shows good tolerance for $t_{InP}$: in fact, increasing the slab thickness by 40 nm leads only to a 20 nm redshift of the cavity mode with small influence on the Purcell factor and maintaining a DCE of ~70% around resonance. This robust behaviour, together with the ability to readjust the resonance wavelength with small variations of $R$ and $a$, is essential for the fabrication of an electrically driven CBG which may require an increased slab thickness to avoid degradation of the device performances due to Zn (p-dopant) diffusion [57,58].

## 4. Conclusion

In conclusion, we presented a detailed study based on FEM simulations of circular Bragg gratings etched in InP slabs and operating in the telecom C-band, which guarantee efficient extraction of photons emitted by InAs/InP QDs and enable the simultaneous enhancement of the X and XX spectral lines.

First, we optimized the choice of the design parameters to achieve the best combination of Purcell factor, bandwidth and collection efficiency: we reported a DCE in NA = 0.65 close to 90% in the whole range 1520 - 1580 nm, together with a cavity mode exhibiting FWHM = 16 nm and Purcell factor up to 15. Then, we determined the optimal strategy for direct coupling of the emitted photons into four different SM fibers. We showed a maximum MCE of 68.5% in UHNA4 and discussed potential modifications to achieve an even higher coupling efficiency of 71.7%. Finally, we proposed a novel device design which is directly compatible with electrical carrier injection, reporting Purcell factors up to 20 and DCE in NA = 0.65 close to 70% in the whole telecom C-band. The absence of particularly demanding features such as very narrow and deep trenches that may be challenging to fabricate is a further benefit of our design compared to previous proposals. Our results may benefit the development of highly efficient electrically driven semiconductor quantum light sources and their integration with existing fiber networks for long-distance quantum communication systems.

**Funding**

This project has received partial funding from the European Union's Horizon 2020 research and innovation programme under the Marie Skłodowska-Curie grant agreement No 721394.

**Acknowledgments**

Andrea Barbiero thanks Aleksander Tartakovskii for academic supervision.